\documentclass[twocolumn,showpacs,preprintnumbers,amsmath,amssymb]{revtex4}

\usepackage{graphicx}
\usepackage{dcolumn}
\usepackage{bm}

\begin{document}

\title{Variation of dispersion measure: evidence of geodetic
precession of binary pulsars}
\author{Biping Gong}

\affiliation{ Department of Astronomy, Nanjing University, Nanjing
210093,PR.China}

\email{bpgong@nju.edu.cn}


\begin{abstract}
Variations of dispersion measure (DM) have been observed in some
binary pulsars, which can not be well explained by the propagation
effects, such as turbulence of the interstellar media (ISM)
between the Earth and the pulsar. This paper provides an
alternative interpretation of the phenomena, the geodetic
precession of the orbit plane of a binary pulsar system. The
dynamic model can naturally avoid the difficulties of propagation
explanations. Moreover the additional time delay represented by
the DM variation of two binary pulsars can be fitted numerically,
through which some interesting parameters of the binary pulsar
system, i.e., the moment of inertia of pulsars can be obtained,
$I_1=(2.0\pm0.6)\times 10^{45}$g~cm$^{2}$. The elimination of the
additional time delay by the dynamic effect means that ISM between
the these pulsars and the Earth might also be stable, like some
other binary pulsars.

\end{abstract}


\maketitle

\section{Introduction}
DM within ISM can delay a radio pulse in reaching Earth by a
number of seconds equal to $DM/(2.41\times 10^{-4}f^{2})$, where
$f$ is the observing frequency in MHz and DM is the column density
of free electrons integrated along the line of sight in unite of
pc cm$^{-3}$\cite{mt77},
\begin{equation}
\label{dmdn} DM=\int_0^{l} n_{e}dz \,,
\end{equation}
where $l$ is the distance to the pulsar. For many pulsars, the DM
can be characterized as a constant that holds steady over years of
observation.

However, millisecond pulsar in the globular cluster 47 Tucanae,
i.e., PSR J0023$-$7203J (47 Tuc J),
shows variations of DM as a function of orbital phase\cite{frei}.
The variations of DM are independent of frequency, which indicates
that the additional time delay is not likely a propagation
effect\cite{frei}, since  propagation effect predicts that the
waves at the low-frequency and high-frequency should show very
different time of arrivals (TOAs).

 The galactic binary pulsar PSR
J0621$+$1002 experiences dramatic variability in its DM\cite{spl}
, with gradients as steep as 0.013 pc cm$^{-3}$ yr$^{-1}$.
If the DM variation is interpreted as spatial fluctuation in the
interstellar electron density, then it would obviously deviate
from the simple power law predicted by the standard theories of
ISM\cite{spl, ric}.

Therefore, as discussed by the authors\cite{frei, spl},
attributing the additional time delay (or residuals of (TOAs) to
propagation effect, DM variation, is not very satisfactory in the
comparison with the observations.

The geodetic precession induced orbital effect of a binary pulsar
system can cause an additional time delay, which can well explain
the long-term variabilities, such as derivatives of the semi-major
axis, $\dot{x}$, $\ddot{x}$, and the orbital period,
$\dot{P}_{b}$, $\ddot{P}_{b}$ measured in PSR~J2051$-$0827 and
PSR~B1957$+$20\cite{go}.

This paper applies the geodetic precession induced time delay in
shorter time scales (relative to secular variabilities) to
interpret the residuals in timing measurement of 47 Tuc J, which
has been attributed to the variation of DM. The new explanation
can fit the residuals and also avoid the frequency difficulty in
the propagation effect.

Moreover, the geodetic precession induced variations at different
time scales impose strong constraints on the intrinsic parameters
of a binary pulsar system. Fitting them together, we can obtain
for the first time numerical result of the spin angular momenta of
the two stars as well as the moment of inertia of the pulsar which
is consistent with theoretical predictions.

The significant DM variation of PSR J0621$+$1002 and PSR
J0024$-$7204H (47 Tuc H) can also be well explained by the dynamic
effect. The elimination of the additional time delay by the
dynamic effect means that the DM (or ISM) of these three binary
pulsars might be  very stable.

The situation is similar to PSR B1855$+$09, which has found no
unexplained perturbation in the long record of TOAs, and lead to a
reduction in the upper limit of energy density in the
gravitational wave background radiation\cite{lob}.

In section II the geodetic precession induced orbital precession
velocity in general case is introduced. In section III the
additional time delay due to geodetic precession of a binary
pulsar system is derived. And in section IV, V and VI the dynamic
effect is applied to  47 Tuc J, PSR J0621$+$1002 and 47 Tuc H
respectively. Section VII summarizes the relation of the geodetic
precession induced effects with the time delay in two cases, and
also  the evidences the geodetic precession in binary pulsars.
\section{orbital precession}

The motion of a binary system can be regarded as the precession of
three vectors, the spin angular momenta of the pulsar and its
companion star, ${\bf S}_1$ and  ${\bf S}_2$, and the orbital
angular momentum $ {\bf L}$. The change of the orbital period due
to the gravitational radiation is  2.5 post-Newtonian order
(2.5\,PPN), whereas  the geodetic precession corresponds to
1.5\,PPN.
So the influence of gravitational radiation on the motion of a
binary system can be ignored when discussing dynamics of a binary
pulsar system. Therefore, the total angular momentum, ${\bf J} =
{\bf L} + {\bf S}_1 + {\bf S}_2$, can be treated as invariable
both in magnitude and direction ($\dot{{\bf J}}=0$). With
$\Omega_0$ denoting the precession rate of $\bf L$ around $\bf J$,
the spin-orbit coupling can be expressed as\cite{bo,apo,kid}
\begin{equation}
\label{e1} {\bf \Omega}_0 \times {\bf L} = -{\bf \Omega}_1
 \times {\bf S}_1 - {\bf \Omega}_2 \times {\bf S}_2\,,
\end{equation}
where $\Omega_1$ and $\Omega_2$ represent the precession of the
pulsar and its companion star, respectively. Ignoring terms over
2\,PPN, $\Omega_{1}$ and $\Omega_{2}$ can be written as\cite{bo}
\begin{equation}
\label{e1aa}\Omega_{1}= \frac {L}{2r^{3}}(4+\frac{3m_{2}}{m_{1}})
\ , \ \ \Omega_{2}= \frac {L}{2r^{3}}(4+\frac{3m_{1}}{m_{2}})
 \ ,
\end{equation}
where $m_1$ and  $m_2$ are masses of the pulsar and the companion
star respectively, and $r$ is the separation of  $m_1$ and  $m_2$.
Notice $L\sim r^{1/2}$, $\Omega_1$ and $\Omega_2$ are 1.5\,PPN.

Barker and O'Connell's two-body equation included two spins, but
the orbital precession velocity was not expressed as relative to
the total angular momentum, ${\bf J}$, therefore, it cannot be
compared to observation directly (${\bf J}$ is static relative to
the line of sight after counting out the proper motion of the
binary system).

Apostolatos et al and Kidder\cite{apo, kid}'s orbital precession
velocity was relative to ${\bf J}$, however their velocity of
orbit plane was derived in the case of one spin,
i.e., ${\bf S}_1=0$. Which is suitable only for special binary
systems, like pulsar-black hole binary.

Therefore, it seems contradictory that in Barker and O'Connell's
equation ${\bf L}$ doesn't precess around ${\bf J}$, but in
practical use, ${\bf L}$ is expressed as precessing around ${\bf
J}$. Actually these two expressions can be consistent in the
scenario which has been mentioned by Smarr and Blandford\cite{sb}.
In which ${\bf L}$, ${\bf S}_1$ and ${\bf S}_2$ all precess around
${\bf J}$ rapidly (1.5PPN), whereas the velocities of ${\bf L}$
relative to ${\bf S}_1$ and ${\bf S}_2$ are very small (2PPN). And
in the confrontation with observation, only the rapid precession
velocity  relative to ${\bf J}$, 1.5PPN, should be used.

 Gong\cite{go} derived the
orbital precession velocity in a general cases, which is relative
to ${\bf J}$, and includes both ${\bf S}_1$ and ${\bf S}_2$. The
derivation is based on two simple assumptions: conservation of the
total angular momentum, Eq($\ref{e1}$), and the geometry
constraints of the triangle formed by ${\bf J}={\bf S}+{\bf L}$
(${\bf S}\equiv{\bf S}_1+{\bf S}_2$). The two assumptions lead to
precession rate of ${\bf L}$ around ${\bf J}$\cite{go},
\begin{equation}
\label{e1a} \Omega_{0}=
\Omega_{2}\sin\lambda_{LS}+
(\Omega_{1}-\Omega_{2})\frac{S^{\parallel}_{1}}{S}\sin
\lambda_{LS_{1}}
 \ \,,
\end{equation}
where $S^{\parallel}_{1}=S_1\cos \eta_{SS1}$, denoting the
component of $\bf{S}_1$ in the plane determined by $\bf{S}$ and
$\bf{J}$. Note that $L\sin \lambda_{LJ}\approx S$ is used in
Eq($\ref{e1a}$), since $S/L\ll 1$.  The right-hand side of
Eq($\ref{e1a}$) can as well be written by replacing subscribes 1
with 2 and 2 with 1.

The geodetic precession of the orbit can cause an additional
apsidal motion. In the case of $S/L\ll 1$, the advance of the
precession of the periastron, $\dot{\omega}$ can be given  by
\cite{sb}
\begin{equation}
\label{e3i}\dot{\omega}^{obs}=\dot{\omega}^{GR}+\Omega_{0}\cos\lambda_{LJ}
\approx\dot{\omega}^{GR}+\Omega_{0} \ \,.
\end{equation}
As given by Eq($\ref{e1a}$), $\Omega_{0}$ can be as large as
$\dot{\omega}^{GR}$ (1.5\,PPN), the GR prediction of the advance
of periastron.

The effect of spin-orbit coupling on secular evolution of the
orbital inclination, $i$, can be given by
\begin{equation}
\label{e2}
 \cos i=\cos \lambda _{LJ} \cos I - \sin \lambda _{LJ} \sin I \cos\eta_0 \ \,,
\end{equation}
where $I$ is the angle between the total angular momentum, ${\bf
J}$, and the line of sight, and $\eta_0=\Omega_{0}t+\eta _{i} $
($\eta_{i} $ is the initial phase) is the phase of precession of
${\bf L}$. Thus $i $ is also a function of time. By
 Eq($\ref{e2}$)
 the first derivative of
the projected semi-major axis is\cite{go}
\begin{equation}
\label{e3c}\dot{x}=-x\Omega_{0}\sin \lambda_{LJ}\sin\eta_0 \cot i
\ \,.
\end{equation}
However, since
 $\bf{S}_1$ and $\bf{S}_2$ precess with different velocities,
$\Omega_{1}$ and $\Omega_{2}$ respectively ($m_1\neq m_2$), then
$\bf{S}$ varies in both magnitude and direction ($\bf{S}_1$,
$\bf{S}_2$ and $\bf{S}$ form a triangle), then from the triangle
of $\bf{S}$, $\bf{L}$ and $\bf{J}$,  in react to the variation of
$\bf{S}$, $\bf{L}$ must vary in direction ($|{\bf L}|=$const),
which means the variation of $\lambda_{LJ}$ ($\bf{J}$ is
invariable).

The change of $\lambda_{LJ}$ means that the orbital plane tilts
back and forth.  In turn, both $\lambda_{LS}$ and $\lambda_{JS}$
vary with time. Therefore, from Eq($\ref{e1a}$), the derivative of
the rate of orbital precession can be given by\cite{go},
\begin{equation}
\label{en1} \dot{\Omega}_{0} =\Omega_{2}\Omega_{12}X_3X_4-
\Omega_{12}X_1(\Omega_{01}X_2+\Omega_{12}X_3) \ \,,
\end{equation}
where $\Omega_{12}=\Omega_{1}-\Omega_{2}$,
$\Omega_{01}=\Omega_{1}-\Omega_{0}$,
 $X_1=\frac{S^{\parallel}_{1}}{S}\sin \lambda_{LS_{1}}$,
 $X_2=\tan\eta_{ss1}$,
 $X_3=\frac{S_{V1}S_{V2}}{S^{2}}\frac{\sin\eta_{s1s2}}{\alpha\sin\lambda_{JS}}
 $, and
 $X_4=\frac{\cos^{2}\lambda_{LS}}{\sin \lambda_{LS}}$,
with $\alpha=\sin \lambda_{JS}+\frac{\cos
^{2}\lambda_{LS}}{\sin\lambda_{LS}}$, $S_{V1}=S_{1}\sin
\lambda_{JS1}$ and $S_{V2}=S_{2}\sin \lambda_{JS2}$ represent
components of $\bf{S}_{1}$ and $\bf{S}_{2}$ that are vertical to
$\bf{J}$.

 Note that $\Omega_1$ and
$\Omega_2$ are unchanged when ignoring the orbital decay (2.5PPN),
and $\lambda_{LS_{\alpha}}$ are unchanged also, since they decay
much slower than the orbital decay\cite{apo}. $\ddot{\Omega}_{0}$
can be easily obtained through Eq($\ref{en1}$).

$\dot{\Omega}_{0}$, the derivative of $\Omega_{0}$, can be
absorbed by $\dot{P}_{b}$. The variation in the precession
velocity of the orbit results in a variation of orbital frequency
(${\nu}_{b}=2\pi/P_{b}$),
${\nu}^{\prime}_{b}-{\nu}_{b}=\dot{\Omega}_{0}\Delta t$ .
Then we have $\dot{\nu}_{b}=\dot{\Omega}_{0}$, therefore\cite{go},
\begin{equation}
\label{ez2} \dot{P}_{b}=-\frac{\dot{\Omega}_{0}P_{b}^{2}}{2\pi} \
\,.
\end{equation}
From Eq($\ref{en1}$) and Eq($\ref{ez2}$), we can see that the
contribution of $\dot{\Omega}_{0}$ to $\dot{P}_{b}$ can be as
large as 1~PPN, which is much larger than the contribution of GR
to $\dot{P}_{b}$ (2.5PPN). While $\dot{P}_{b}$ can also be much
smaller than 1~PPN in special combination of parameters in
Eq($\ref{en1}$).

\section{geodetic precession induced time delay}
As discussed in section II, the geodetic precession induced
orbital effect results an additional apsidal motion which can be
absorbed by the post-Kepler parameter, $\omega^{obs}$, an
additional precession of orbital plane which can be absorbed by,
$\dot{x}$, and an additional variation of orbital period which can
be absorbed by $\dot{P}_{b}$\cite{go}.

These additional effects of a binary system can not only cause
long-term (secular) time delay, but also short-term time delay.

The essential transformation relating solar system barycentric
time $t_b$ to pulsar proper time $T$ is summarized by the
expression\cite{tw}
\begin{equation}
\label{tw1} t_b-t_0=T+\Delta_{R}+\Delta_{E}+\Delta_{S}+\Delta_{A}
 \,,
\end{equation}
where $\Delta_{R}$ is the "Roemer time delay", is the propagation
time across the binary orbit; $\Delta_{E}$ and $\Delta_{S}$ are
the orbital Einstein and Shapiro delays; and $\Delta_{A}$ is a
time delay related with aberration caused by rotation of the
pulsar. The dominant time delay, $\Delta_{R}$ is given\cite{tw}
\begin{equation}
\label{tw2} \Delta_{R}=xF(\omega+u)
 \,,
\end{equation}
where
$$
F(\omega+u)=\sin \omega [\cos u-e(1+\delta_r)]+
$$
\begin{equation}
\label{tw2a} [1-e^{2}(1-\delta_{\theta})^{2}]^{1/2}\cos \omega\sin
u
 \,.
\end{equation}
In calculation the small quantities, $\delta_{r}$ and
$\delta_{\theta}$ due to aberration are ignored. $u$ and
$A_{e}(u)$ are the eccentric anomaly and the true anomaly
respectively. The relations of  $u$, $A_{e}(u)$ and the longitude
of periastron, $\omega$ are given\cite{tw}
\begin{equation}
\label{tw3} u-e\sin
u=2\pi[(\frac{T-T_0}{P_b})-\frac{\dot{P}_{b}}{2}(\frac{T-T_0}{P_b})^{2}]
\,,
\end{equation}
\begin{equation}
\label{tw4} A_{e}(u)=2\arctan[(\frac{1+e}{1-e})^{1/2}\tan
\frac{u}{2}]
 \,,
\end{equation}
\begin{equation}
\label{tw5} \omega=\omega_0+kA_{e}(u)
 \,,
\end{equation}
where $k\equiv \dot{\omega}P_{b}/(2\pi)$. Eq($\ref{tw1}$) to
Eq($\ref{tw5}$) are the standard treatment in pulsar timing
measurement. When consider the contribution of orbital precession
to the time of arrival, the Roemer delay  should has a different
value, $\Delta_{R}^{\prime}$, relative to the standard,
 $\Delta_{R}$, which doesn't include the dynamic effect. The deviation leads to an
additional time delay.
\begin{equation}
\label{opt2}
\delta\Delta_{R}=\Delta_{R}^{\prime}-\Delta_{R}=x^{\prime}F(\omega^{\prime}+u^{\prime})-
xF(\omega+u)
 \,.
\end{equation}
$x^{\prime}=x+\dot{x}t$ at right-hand side of Eq($\ref{opt2}$)
corresponds to an additional time delay by the precession of the
orbital plane, in which $\dot{x}$ is given by Eq($\ref{e3c}$).
 $u^{\prime}$ represents an
additional time delay by the nutation, in which $\dot{P}_{b}$ of
Eq($\ref{tw3}$)) should be given by Eq($\ref{ez2}$)).  And
$\omega^{\prime}$ represents an additional time delay by the
apsidal motion, in which $k$ of Eq($\ref{tw5}$) should be written
as $k^{\prime}= (\dot{\omega}+\Omega_{0})P_{b}/(2\pi)$.

Thus $\delta\Delta_{R}$ of Eq($\ref{opt2}$) at two different
moment, $t$ and $t+\tau$ can be given as
$$\delta\Delta_{R}^{t+\tau}-\delta\Delta_{R}^{t}=
[x^{\prime}F(\omega^{\prime}+u^{\prime})- xF(\omega+u)]_{t+\tau}
$$
\begin{equation}
\label{opt3}-[x^{\prime}F(\omega^{\prime}+u^{\prime})-
xF(\omega+u)]_{t}
 \,.
\end{equation}

In other words if the true Roemer delay is given by
$\Delta_{R}^{\prime}$, which include the dynamic effect, but it is
treated as the standard one $\Delta_{R}$ without the dynamic
effect, then additional time delay results. Following section
shows that the additional time delay can be well fitted by the
difference between $\Delta_{R}^{\prime}$ and $\Delta_{R}$ given by
Eq($\ref{opt2}$) and  Eq($\ref{opt3}$).

\section{47 Tuc J}
\subsection{estimations}


This section shows that in the geodetic precession model, once an
orbital precession velocity, $\Omega_0$, which is suitable to
explain one secular variability, i.e., $\dot{x}$, then it is also
suitable to explain all the other secular variabilities, like
$\dot{\omega}$ and $\dot{P}_{b}$ and $\dot{DM}$ of a binary
pulsar. Which indicate that the physics underlying these phenomena
is most likely the geodetic precession.

With $m_1=1.44M_{\odot}$, $m_2=0.02M_{\odot}$, ${P}_{b}=0.12$d and
$x^{obs}=0.04$\cite{frei}, we have the semi-major axis,
$a=[GM/\nu_{b}^{2}]^{1/3}=8.2\times 10^{10}$~cm, and orbital
angular momentum, $L=\nu_{b}\mu a^{2}(1-e^{2})^{1/2}=1.6\times
10^{50}$g~cm$^{2}$s$^{-1}$, with $\mu$ the reduced mass. Then we
have $\Omega_1=4.4\times 10^{-11}$s$^{-1}$, $\Omega_2=2.5\times
10^{-9}$s$^{-1}$ by Eq($\ref{e1aa}$). Assume
$\Omega_0\approx\Omega_1$, and by Eq($\ref{opt2}$) the geodetic
precession induced time delay in the time interval $\tau=1$yr is
approximately,
\begin{equation}
\label{tucj} |\delta\Delta_{R}|\approx
x|\delta\sin(\omega)|\approx x\Omega_0\tau|\cos
(\omega)|\approx5.5\times 10^{-7} (s)
 \,,
\end{equation}
while in the case $\Omega_0\approx\Omega_2$, the additional time
delay in one year is  $|\Delta_{R}|\approx 3.2\times 10^{-5} $.
The observed one is $\dot{DM}<2\times
10^{-4}$cm$^{-3}$pc/yr\cite{frei}, which corresponds to $2\times
10^{-6}$ s per yr.

Therefore, if $\Omega_1<\Omega_0<\Omega_2$, i.e.,
$\Omega_0\approx\Omega_2/10$, then the corresponding time delay,
$|\Delta_{R}|\approx 3.2\times 10^{-6} $s, is close to the
observational limit.

The geodetic precession induced secular variabilities, $\dot{x}$
and $\dot{P}_{b}$ can also be estimated. With
$\Omega_0\approx\Omega_1$  and $\dot{x}^{obs}=(-2.7\pm 0.7)\times
10^{-14}$\cite{frei},
 Eq($\ref{e3c}$) becomes
\begin{equation}
\label{tucj2}
 1.8\times10^{-12}\sin \lambda_{LJ}|\sin\eta_0\cot i|\approx2.7\times 10^{-14}  \,.
\end{equation}
Thus $\sin\lambda_{LJ}|\sin\eta_0\cot i|\approx 1.5\times10^{-2}$.
  If $|\sin\eta_0\cot i| \approx 0.1$, then $\lambda_{LJ}\approx
  1.5\times10^{-1}$, which means
 $S/L\approx 1.5\times10^{-1}$. With $L$ obtained above, we have $S_1\approx S\approx
 2.4
\times10^{49}$g~cm$^{2}$s$^{-1}$. Having the measured pulsar
period, $P=2.1$ms, the moment of inertia of the pulsar is
$I_{1}\approx 8.0I_{45}$ ($I_{45}\equiv
1\times10^{45}$g~cm$^{2}$). While if assuming
$\Omega_0\approx\Omega_2$, then the moment of inertia is
$I_{1}\approx 0.14I_{45}$. And similarly in the case
$\Omega_0\approx\Omega_2/10$, $I_{1}\approx 1.4I_{45}$. Which
means the velocity that suitable for $\dot{DM}$ corresponds to a
moment of inertia that is very close to the theoretical
prediction.

As shown in  Eq($\ref{ez2}$), $\dot{P}_b$ is determined by
$\dot{\Omega}_{0}$, which can vary in a range of several order of
magnitude by different combination of variables in
Eq($\ref{en1}$). If we assume
$\dot{\Omega}_{0}\approx\Omega_{1}^{2}\approx 1.9\times
10^{-21}$s$^{-2}$, then by Eq($\ref{ez2}$), we have
$\dot{P}_{b}\approx 0.33\times10^{-13}$. Which is about one order
of magnitude smaller than the measured one.
$\dot{P}_{b}^{obs}=(-0.5\pm0.13)\times 10^{-12}$.
Whereas,  if we assume $\dot{\Omega}_{0}\approx\Omega_{2}^{2}$,
then $\dot{P}_{b}\approx 9.8\times10^{-11}$. Which is two order of
magnitude larger than $\dot{P}_{b}^{obs}$.

And in the case $\Omega_0\approx\Omega_2/10$, $\dot{P}_{b}\approx
9.8\times10^{-13}$, which is close to the measured
$\dot{P}_{b}^{obs}$.

Therefore, the orbital velocity of order of magnitude,
$\Omega_0\approx\Omega_2/10$, can consistent with measured
variabilities. Which indicates that the geodetic precession might
the true mechanism that responsible for the observational results.

\subsection{fitting}
DM variation with the orbital phase is clearly detected in 47 Tuc
J, as shown by the scattered points with error bars in Fig~2.
The variation has been interpreted as a cometary-like phenomenon,
which caused by material at a considerable distance from the
companion, with much higher average electron density than that of
the rest area. In other words, the plasma cloud is responsible for
the additional time delay, which can be attributed to the
variations of DM .

Whereas the variation of DM and the residuals of TOAs as function
of orbital phase are very close at different frequencies, from
660MHz to 1486MHz\cite{frei}. Which means that the additional time
delay is not likely caused by the propagation effect of ISM. Since
the time delay due to DM variations should be very different at
low and high-frequency.

Actually, what measured in 47 Tuc J of  Fig~2 is residual of TOAs,
or additional time delay, $\Delta^{obs}$, which can be attributed
to the variation of DM. And the discussion above indicates that
such explanation is difficult to explain the frequency problem.

Therefore, we can transform the DM variation of Fig~2 back into
the additional time delay, $\Delta^{obs}$, and try to interpret it
by the dynamic effect.

The time delay, $ \Delta^{obs}$, measured in the time interval
between two moment, $t$ and $t+\tau$ (or orbital phases $\phi(t)$
and $\phi(t+\tau)$), corresponds to the DM variation, $DM(t)$ and
$DM(t+\tau)$ respectively. The relationship between DM and $
\Delta^{obs}$ can be given by\cite{mt77}
\begin{equation}
 \label{dmt1}
\delta\Delta^{obs}=\frac{DM(t+\tau)}{2.41\times 10^{-4}
f^{2}}-\frac{DM(t)}{2.41\times 10^{-4}f^{2}} \, .
\end{equation}
Thus the geodetic precession induced time delay at two moment $t$
and $t+\tau$ given by Eq($\ref{opt3}$),  can be used to explain
the observational one represented by Eq($\ref{dmt1}$), $
\delta\Delta^{obs}$. The new interpretation can not only solve the
difficulties in the previous explanations of 47 Tuc J, but also
fit the additional time delay numerically.

The estimation above indicates that  $\delta\Delta_{R}$, $\dot{x}$
and  $\dot{P}_{b}$ can be well explained by the geodetic
precession of the binary system. Now we can fit these variations
numerically.

The vectors ${\bf S}_1$, ${\bf S}_2$ and ${\bf S}$ are studied in
the coordinate system of the total angular momentum, in which the
z-axis directs to ${\bf J}$, and the x- and y-axes are in the
invariance plane. ${\bf S}$ can be represented by $S_{P}$ and
$S_{V}$, the components parallel and vertical to the z-axis,
respectively:
\begin{equation}
\label{ex1}
 S=(S_{V}+S_{P})^{1/2} \ .
\end{equation}
$S_{P}$ and $S_{V}$ can be expressed (recall $S_{V1}$, $S_{V2}$
and $S_{V}$ form a triangle) as
$$S_{P}=S_{1}\cos \lambda_{JS1}+S_{2}\cos \lambda_{JS2} \ ,$$
\begin{equation}
\label{ex2}  S_{V}= (S_{V1}^{2}+S_{V2}^{2}-2S_{V1}S_{V2}\cos
\eta_{S1S2})^{1/2} \ ,
\end{equation}
where $\eta_{S1S2}$ is the misalignment angle between $S_{V1}$ and
$S_{V2}$, which can be written as
\begin{equation}
\label{ex5} \eta_{S1S2}=(\Omega_1-\Omega_2)t+\phi_{i}  \ .
\end{equation}
Therefore, by the variation of $S$ as function of time (in the
case of one spin, $S=$const), we can obtain $\Omega_0$ as function
of time through Eq($\ref{e1a}$). Thus the measured $\Delta^{obs}$
of Eq($\ref{dmt1}$) (or $\dot{DM}$) can be fitted step by step
through Eq($\ref{opt3}$).
\begin{equation}
\label{ex5a}
\delta\Delta^{obs}=\delta\Delta_{R}^{t+\tau}-\delta\Delta_{R}^{t}
\ .
\end{equation}
Through Eq($\ref{ez2}$) and Eq($\ref{e3c}$), the measured secular
variabilities,  $\dot{P}_{b}^{obs}$ and
$\dot{x}^{obs}$\cite{frei}, can also be fitted along with
$\Delta^{obs}$  by Monte-Carlo method. Obviously the long-term and
short-term together imposes very stringent constraints on the
numerical solutions. Notice that the measured mass function,
$f_1(M_{\odot})=4.864\times 10^{-6}$\cite{frei} is also considered
in the fitting.

As shown in Eq($\ref{ez2}$) and Eq($\ref{e3c}$), $\Omega_0$ is
included in $\delta\Delta_{R}$ and $\dot{x}$, meanwhile
$\dot{\Omega}_0$ is included in $\dot{P}_{b}$. Both  $\Omega_0$
and $\dot{\Omega}_0$ contain  $S(t)$, $S_1$ and angles, as shown
in Eq($\ref{e1a}$) and Eq($\ref{en1}$). Therefore, fitting the
short-term and long-term parameters lead to the determination of
$S_1$ and $S_2$, and in turn the moment of inertia of the pulsar,
$I_1$, since  the pulsar period is known.

In numerical fitting, $S_1$ and $S_2$ are fitted in the range
$[2.5\times 10^{46}, 1.4\times 10^{49}]$ and $[2.5\times 10^{46},
7.6\times 10^{49}]$ (g~cm$^{2}$s$^{-1}$) respectively, which are
enough to cover the estimated values. The best solution is shown
in Table~I. By Eq($\ref{dmt1}$) and Eq($\ref{e1a}$), we have
$\dot{DM}\propto \delta\Delta_{R}\propto \Omega_0 \propto S_1$,
thus the $30\%$ errors in the DM variation in the upper plot of
Fig~2, can cause about $30\%$ error in the fitted results, such as
$S_1$ and $I_1$.


After fitting the measured time delay by the predicted one, as
Eq($\ref{ex5a}$), we can transform the predicted one, at the
right hand side of Eq($\ref{ex5a}$) into the theoretical DM
variation, as displayed by the solid curves of Fig 2. So that it
can be compared with the measured ones more clearly.

The measured DM variation are slightly different at different
time, i.e., 1998 June and 1999 October, as shown in Fig~2 (points
with error bars), which can be explained by Eq($\ref{opt2}$) from
which $\delta\Delta_{R}$ are slightly different at different time.


\section{PSR~J0621$+$1002}
 For PSR~J0621$+$1002, dramatic variability
of its DM  has been measured, with gradient as steep as 0.013 pc
cm$^{-3}$yr$^{-1}$, shown in Fig~3. By the standard picture, the
turbulence spreads energy from longer to shorter length scales
arises a power law of the structure function, which is given by
$D_{DM}(\tau)\equiv \langle [DM(t+\tau)-DM(t)]^2 \rangle$, where
$\tau$ is the time lag between DM measurement. However the
structure function obtained from observation obviously deviates
from the simple power law\cite{spl}. Moreover, there is also no
obvious differences in DM variation corresponding to 430 and 1410
MHz\cite{spl}. Which indicates that DM variation of this binary
pulsar is also independent of frequency. In the geodetic
precession induced model these difficulties can be explained
naturally.

Similarly, with $m_1=1.70M_{\odot}$, $m_2=0.97M_{\odot}$,
$e=0.00246$, ${P}_{b}=8.3$d and $x=12$\cite{frei}, we have
$\Omega_1=1.4\times 10^{-12}$s$^{-1}$, $\Omega_2=2.2\times
10^{-12}$s$^{-1}$. Assuming $\Omega_0\approx\Omega_2$, the
geodetic precession induced time delay given by Eq($\ref{tucj}$)
can be written as ($\tau=1$yr)
\begin{equation}
\label{0621} |\delta\Delta_{R}|\approx x\Omega_0\tau|\cos
(\omega)|\approx 8.3\times 10^{-4} (s)
 \,,
\end{equation}
The measured maximum DM variation, $\dot{DM}=0.013$pc
cm$^{-3}$yr$^{-1}$\cite{spl}, corresponds to a delay of
$\Delta^{obs}=2.9\times 10^{-4}$s per year. Therefore the
predicted one and the measured one can be consistent.

\section{47 Tuc H}
The DM variation of 47 Tuc H,
$\dot{DM}=-0.024(3)$cm$^{-3}$pc~yr$^{-1}$ (equivalent
$\Delta_{R}=5.4\times 10^{-4}$s per year), is one of the largest
DM variations for any pulsar, and no binary parameters available
can explain the trend\cite{frei}. Whereas the geodetic precession
model can explain this large DM variation easily.

By the same treatment as the two binaries above, with
$m_1=1.50M_{\odot}$, $m_2=0.186M_{\odot}$
and  $x^{obs}=2.15$\cite{frei},  the additional time delay per
year is approximately, $\Delta_{R}\approx 1.8\times 10^{-4}$s, in
the case $\Omega_0\approx\Omega_1\approx
2.7\times10^{-12}$s$^{-1}$; and $\Delta_{R}\approx 1.2\times
10^{-3}$s, in the case $\Omega_0\approx\Omega_2\approx
1.7\times10^{-11}$s$^{-1}$. Therefore, the large $\dot{DM}$
measured in 47 Tuc H, which is $\Delta_{R}=5.4\times 10^{-4}$s per
year, can be well explained by an orbital precession velocity that
is in the range, $\Omega_1<\Omega_0<\Omega_2$.

\section{discussion}
As shown in Fig~4, the the orbital precession velocity,
$\Omega_0$, can contribute to the orbital precession, absorbed by
$\dot{x}$; apsidal motion, absorbed by $\dot{\omega}$; and
nutation, absorbed by $\dot{P}_{b}$ respectively.

For very special NS--NS or NS--WD binary pulsars, $m_1=m_2$ or one
spin is ignorable (i.e, $S_1=0$),  ${\bf S}$ is a constant vector,
then $\Omega_0$ is a constant, which means $\dot{\Omega}_0=0$, and
in turn $\dot{P}_{b}=0$, therefore, there will be only static
orbital precession and apsidal motion but no nutation. In such
special cases the additional time delay can be absorbed by
$\dot{x}$ and $\dot{\omega}$, of which $\dot{x}$ is a function of
time (or orbital phase) and $\dot{\omega}$ is unchanged.

While for  general NS--NS or NS--WD binary pulsars, $m_1\neq m_2$
and $S_1\neq 0$, $S_2\neq 0$, the two spins precesses at different
velocities ($\Omega_1\neq \Omega_2$), therefore, ${\bf S}$ varies
both in direction and magnitude (${\bf S}={\bf S}_1+{\bf S}_2$).
Then $\Omega_0$ is a function of time by  Eq($\ref{e1a}$)
($\dot{\Omega}_0\neq 0$), which leads to the nutation
($\dot{P}_{b}\neq 0$) by Eq($\ref{ez2}$). Therefore, for a general
NS--NS or NS--WD binary pulsar, there is nutation effect beside
the precession of the orbit and the apsidal motion.

Thus, for a general binary pulsar, the three constant parameters
$\dot{x}^{obs}$, $\dot{\omega}^{obs}$, and $\dot{P}_{b}^{obs}$
together, can largely eliminate the trend of residuals, or the
additional time delay which can be represented as $DM$ vs time, or
$DM$ vs orbital phase.

Since $\Omega_0$ varies with time, then $\dot{x}$, $\dot{\omega}$
and $\dot{P}_{b}$ also vary with time, as shown by
Eq($\ref{e3c}$), Eq($\ref{e3i}$) and Eq($\ref{ez2}$) respectively,
whereas, the measured $\dot{x}^{obs}$, $\dot{\omega}^{obs}$ and
$\dot{P}_{b}^{obs}$, are all constants, or  the average values of
the true effects, thus for such binary pulsar which have small
orbital periods, i.e., a few hours ($\dot{x}^{obs}$ and
$\dot{P}_{b}^{obs}$ vary rapidly), higher order derivatives, such
as $\ddot{x}^{obs}$ and $\ddot{P}_{b}^{obs}$ are necessary to
eliminate the trend of residuals. The fitting of the $DM$ vs
orbital phase in 47 Tuc J in this paper actually included such
higher order of derivatives, because $\dot{x}$, $\dot{\omega}$,
and $\dot{P}_{b}$ used in fitting are given by Eq($\ref{e3c}$),
Eq($\ref{e3i}$) and Eq($\ref{ez2}$) respectively, which are all
functions of time.

The relationship of the geodetic precession induced secular
variabilities and the additional Roemer time delay as well as the
DM variation is summarized in Fig~3.

Therefore, comparing with the secular variations, $\dot{x}$ and
$\dot{P}_{b}$,  the variation of DM of 47 Tuc J is just the
short-term effect ($t\approx P_b/20$) of geodetic precession in a
binary pulsar.

$\dot{DM}$, and secular variabilities, such as $\dot{x}$ and
$\dot{P}_{b}$, have been  interpreted separately by different
models. While the geodetic precession provides an unified model
which can well explain these variabilities (both short term and
long-term).

On the other hand, the DM variations provide new evidences of the
geodetic precession effect in binary pulsar systems, beside the
secular variabilities, and the theoretical prediction of the
moment of inertia of neutron stars.

The numerical results of spin angular moment of white dwarf star
and  the pulsar, as well as moment of inertia of the pulsar
provides very useful information on both the structure of  white
dwarf stars and neutron stars.

The interpretation of the DM variation by the dynamic effects in
the three binary pulsars (the frequency problems in the previous
explanation are solved automatically) indicts that the  structure
of the ISM between the these pulsars and the  Earth might be very
stable, which is similar as PSR B1855+09\cite{lob}. This might
provide new information in the understanding of ISM.


\begin{table}
\begin{center}
\caption{Parameters obtained by fitting DM vs time,
${\dot{x}}^{obs}$ and ${\dot{P}_{b}}^{obs}$ of 47 Tuc J}
\begin{tabular}{lllllllll}
\hline \hline   $\lambda_{JS_{1}}$ & $\lambda_{JS_{2}}$ &
$\lambda_{LS_{1}}$ &  $\phi_i$ & $\omega_{i}$ & $\eta_i$\\
$0.9192$ & $1.476$ &
 $0.9203$ & $0.1095$ & $3.594$ & $0.3640$ \\\hline

$S_1$(g~cm$^{2}$s$^{-1}$) & $S_2$(g~cm$^{2}$s$^{-1}$) & $m_1$ &
$m_2$ & $e$ & $i$
 \\
 $5.985\times 10^{48}$ & $4.933\times 10^{49}$
& $1.949$ & $0.0264$ & $1.929\times 10^{-5}$ & $1.570$
\\ \hline\hline
\end{tabular}
\end{center}
{\small By $S_1$ the moment of inertia of the pulsar can be
obtained, $I_1=2.0\times 10^{45}$g~cm$^{2}$. Notice that $m_1$ and
$m_2$ are in sun mass, all angles are in radian. The errors of the
values in the above table is about $30\%$
 }
\end{table}

\begin{figure}[t]
\begin{center}
\includegraphics[87,87][700,700]{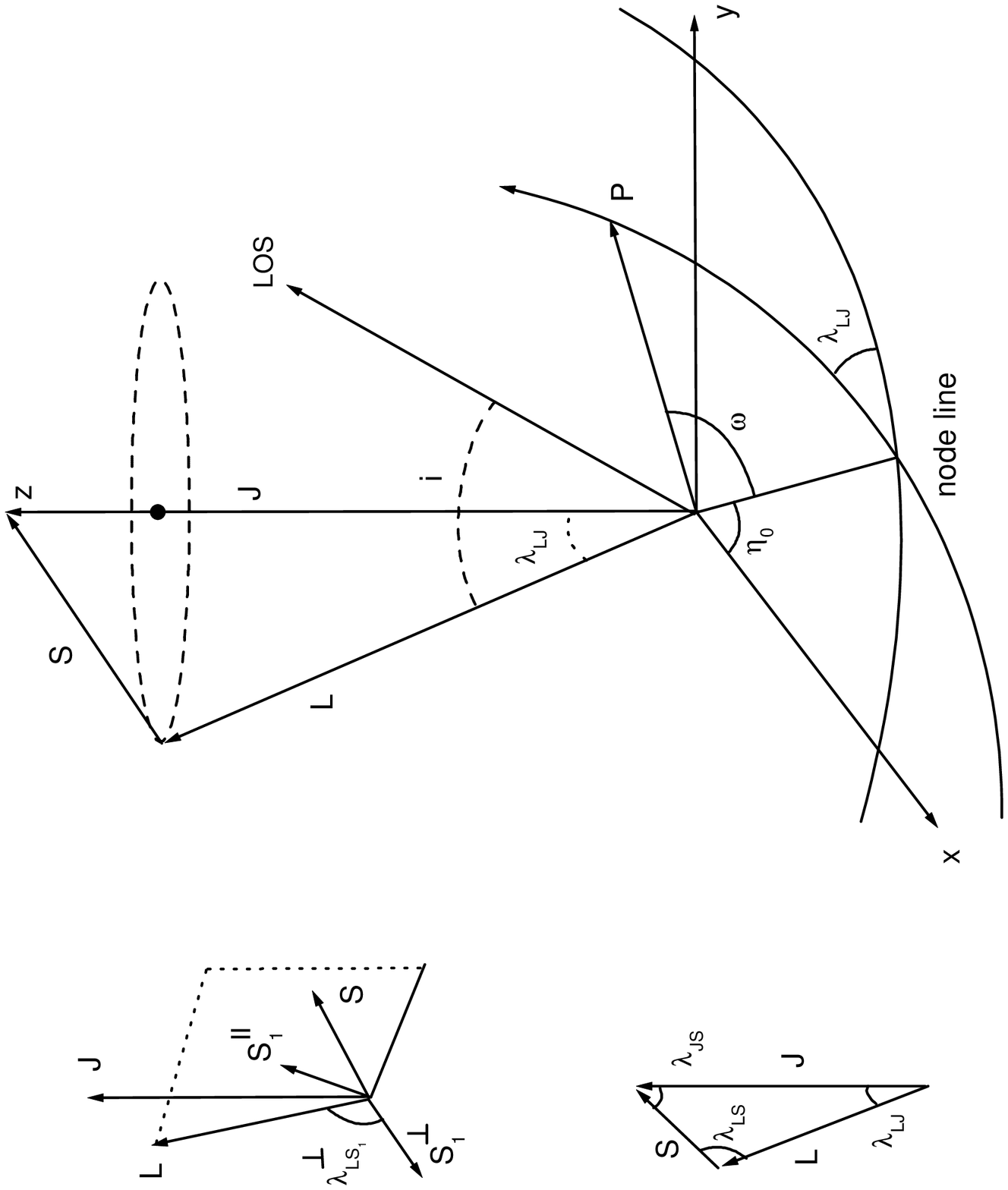}
\end{center}
\caption{Binary geometry and definitions of angles. The invariable
plane (x-y) is perpendicular to the total angular momentum, ${\bf
J}$. The inclination of the orbital plane with respect to the
invariable plane is $\lambda_{LJ}$, which is also the precession
cone angle of  ${\bf L}$ around ${\bf J}$.   The orbital
inclination with respect to the line of sight is $i$. $\eta_0$ is
the phase of the orbital plane precession. $\omega$ is longitude
of periastron (point P). }
\end{figure}

\begin{figure}[t]
\begin{center}
\includegraphics[87,87][700,700]{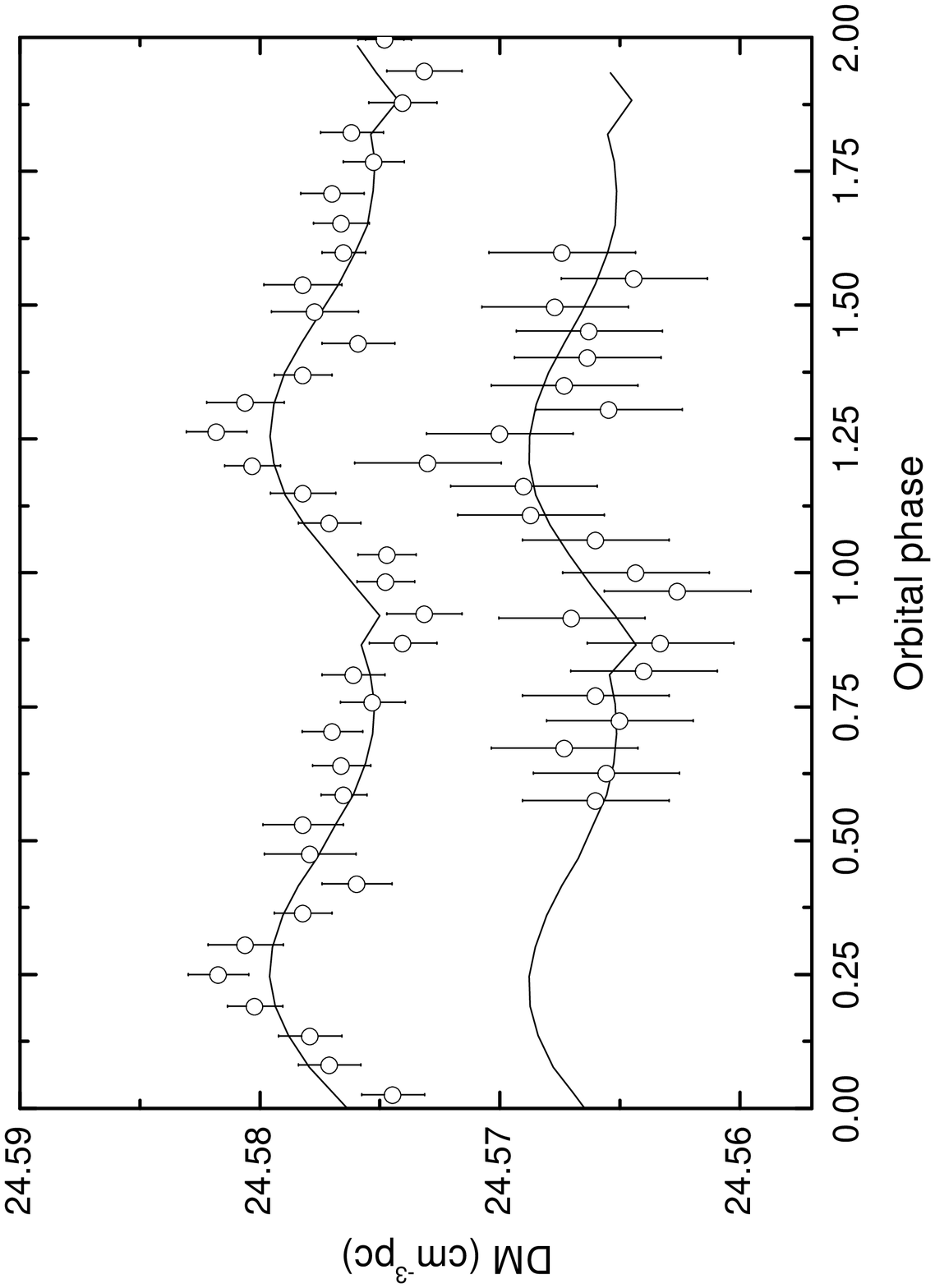}
\end{center}
\caption{The measured\cite{frei} and predicted DM as a function of
orbital phase in 47 Tuc J . The measured upper plot represents the
measurement made with 1999-2002 high-resolution data, displayed
$0.01$cm$^{-3}$ pc below their measured values for clarity. The
second plot represent the DM obtained using only the data from the
best observation (1999 October 11), displayed $0.02$cm$^{-3}$ pc
below their measured values\cite{frei}. The solid curve of the
upper plot represent the best fit. And the solid curve of the
second plot is obtained by setting the time 0.3714 year later than
the upper fitting.
}
\end{figure}

\begin{figure}[t]
\begin{center}
\includegraphics[87,87][700,700]{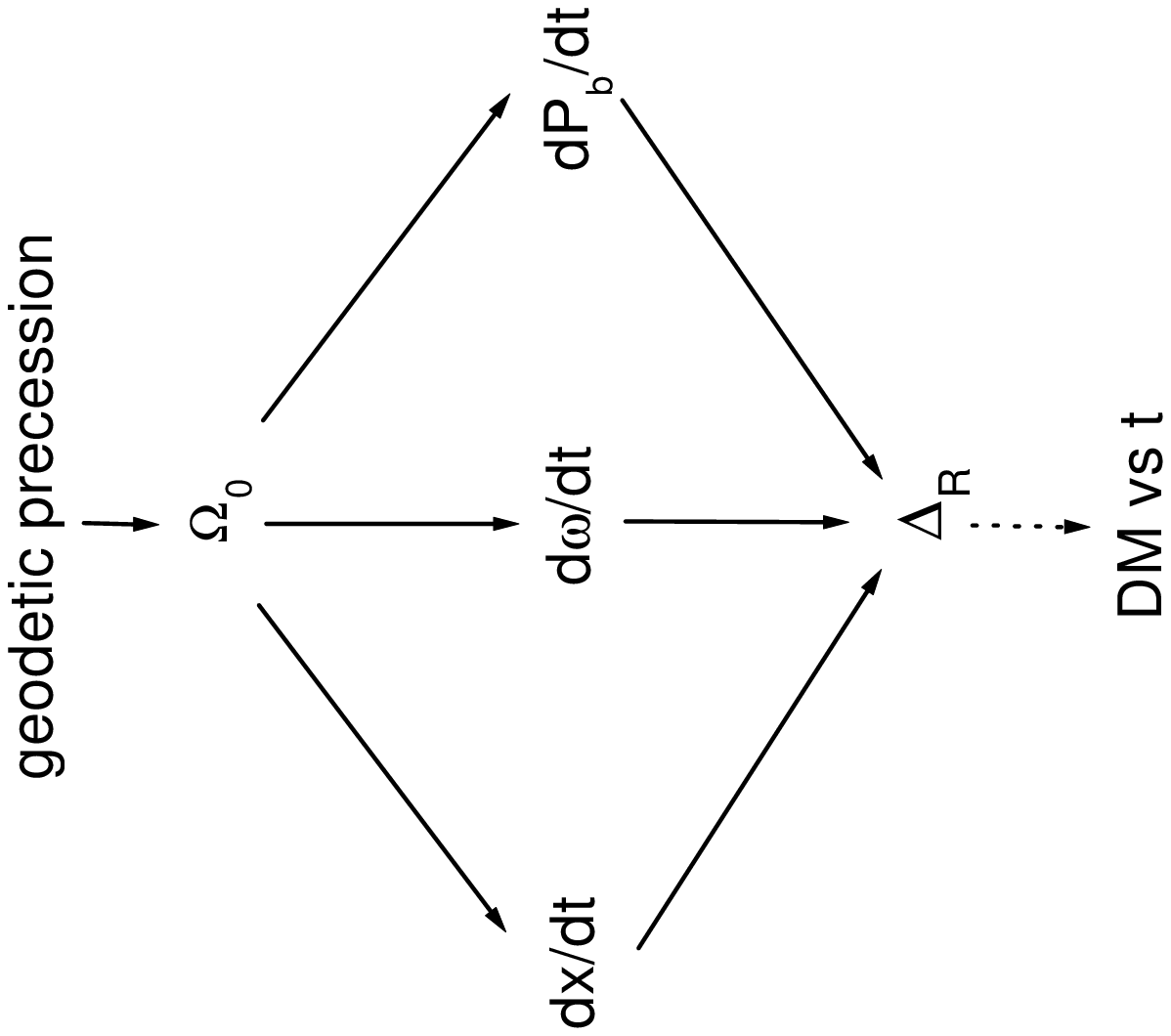}
\end{center}
\caption {Geodetic precession leads to an additional orbital
precession velocity, $\Omega_0$, which can be absorbed by three
parameters, $\dot{x}$, $\dot{\omega}$ and $\dot{P}_{b}$. Which in
turn cause an additional time delay, $\Delta_{R}$, which can be
attributed to DM variation.}
\end{figure}

\end{document}